%             G. 't Hooft  macros version Oct.96  
\newread\testifexists  
\def\GetIfExists #1 {\immediate\openin\testifexists=#1  \ifeof\testifexists\immediate\closein\testifexists\else   
\immediate\closein\testifexists\input #1\fi}    
 \def\epsffile#1{Figure: #1} 					%%% 
\immediate\openin\testifexists=e  \ifeof\testifexists\immediate\closein\testifexists\else   
\immediate\closein\testifexists\input e\fipsf  
 \magnification= \magstep1	% or use \magstephalf    
\tolerance=1600  
\parskip=7pt 
 \baselineskip= 5 true mm \mathsurround=1pt  
\font\smallrm=cmr8

\font\medrm=cmr9

\font\bigbf=cmbx12 
\def\Bbb#1{\setbox0=\hbox{$\tt #1$}  \copy0\kern-\wd0\kern .1em\copy0} 
\immediate\openin\testifexists=a  \ifeof\testifexists\immediate\closein\testifexists\else   
\immediate\closein\testifexists\input a\fimssym.def %% for \Bbb A - Z %% 
\def\secbreak{\vskip8pt plus 1in \penalty-200\vskip 0pt plus -.8in}  
 %\prefbreak{distance}  
\def\hugeskip{\vskip12mm plus 3mm}  
\def\Narrower{\par\narrower\noindent}	% never again use TeX's awful \narrower 
\def\Endnarrower{\par\leftskip=0pt \rightskip=0pt}  
	\def\ra{\rightarrow} 
                  
\def\d{\delta}          \def\D{\Delta}  \def\e{\varepsilon} 
             \def\k{\kappa}  \def\l{\lambda}          
 \def\m{\mu}             \def\f{\phi}                 
 \def\n{\nu}                  
 \def\r{\varrho}         \def\s{\sigma}

 \def\HH{{\cal H}}   \def\NN{{\cal N}}  
  
\def\cl{\centerline}      
\def\ni{\noindent}      \def\pa{\partial}       \def\dd{{\rm d}}  
\def\tl{\tilde}                 \def\bra{\langle}       \def\ket{\rangle}  
\def\fn#1{\ifcase\noteno\def\fnchr{*}\or\def\fnchr{\dagger}\or\def 
\fnchr{\ddagger}\or\def\fnchr{\rm\S}\or\def\fnchr{\|}\or\def 
\fnchr{\rm\P}\fi\footnote{$^{\fnchr}$} 
{\scrunch#1\toe}\ifnum\noteno>5\global\advance\noteno by-6\fi 
\global\advance\noteno by 1} 
\def\scrunch{\baselineskip=11 pt \medrm} 
\def\toe{\vphantom{$p_\big($}} 
\newcount\noteno  
% 	 footnote with alternating symbol	%  

\def\fract#1#2{\raise .35 em\hbox{$\scriptstyle#1$}\kern-.25em/ 
\kern-.2em\lower .22 em \hbox{$\scriptstyle#2$}}  
\def\half{\fract12}   
\def\part#1#2{{\partial#1\over\partial#2}} 
 \def\ref#1{${\vphantom{)}}^#1$} 
  
\def\bbf#1{\setbox0=\hbox{$#1$} \kern-.025em\copy0\kern-\wd0  
\kern.05em\copy0\kern-\wd0 \kern-.025em\raise.0433em\box0}  
% boldface in math mode. 
 \def\qu{\ {\buildrel {\displaystyle ?} \over =}\ } 
 \def\df{\ {\buildrel{\rm def}\over{=}}\ }

 \def\ref#1{${\,}^{\hbox{\smallrm #1}}$} 
 
\def\Gbar{\raise.10em\hbox{--}\kern-.35em G}  
  \def\lap{\setbox0=\hbox{$<$}\,\raise .25em\copy0\kern-\wd0\lower.25em\hbox{$\sim$}\,} 
\def\glt{\setbox0=\hbox{$>$}\,\raise .25em\copy0\kern-\wd0\lower.25em\hbox{$<$}\,}  
\def\gap{\setbox0=\hbox{$>$}\,\raise  
.25em\copy0\kern-\wd0\lower.25em\hbox{$\sim$}\,}     
 
\def\in{{\rm in}} \def\out{{\rm out}} \def\Pl{{\rm Pl}} 
\def\Or{\qquad\hbox{or}\qquad} 
  %{\nopagenumbers %     
 {\ }\vglue 1truecm  
 \rightline{THU-98/22}  
 
\rightline{gr-qc/9805079}
 \hugeskip 
 \cl{\bigbf TRANSPLANCKIAN PARTICLES}\medskip  
  \cl{\bigbf AND THE QUANTIZATION OF TIME}                 
 
  \hugeskip  
 \cl{Gerard 't Hooft } 
 \bigskip  
  \cl{Institute for Theoretical Physics}  
 \cl{University of Utrecht, Princetonplein 5}  
 \cl{3584 CC Utrecht, the Netherlands}  
\smallskip\cl{e-mail:\ \ \tt g.thooft@fys.ruu.nl} 
\hugeskip  
 
\ni{\bf Abstract}\Narrower  
 
Trans-Planckian particles are elementary particles accelerated such that their 
energies surpass the Planck value $ \sqrt{\hbar c^5/8\pi G} $. There are several 
reasons to believe that trans-Planckian particles do not represent independent 
degrees of freedom in Hilbert space, but they are controlled by the cis-Planckian 
particles. A way to learn more about the mechanisms at work here, is to study 
black hole horizons, starting from the scattering matrix Ansatz. 
 
By compactifying one of the three physical spacial dimensions, the scattering 
matrix Ansatz can be exploited more efficiently than before.  The algebra of 
operators on a black hole horizon allows for a few distinct representations. 
It is found that this horizon can be seen as being built up from string bits 
with unit lengths, each of which being described by a representation of the $ 
SO(2,1) $ Lorentz group.  We then demonstrate how the holographic principle 
works for this case, by constructing the operators corresponding to a field $ 
\phi(\tilde x,t) $.  The parameter $ t $ turns out to be quantized in units $ 
t_{\rm Planck}/R $, where $ R $ is the period of the compactified dimension. 
\Endnarrower 
 
\secbreak{\ni\bf 1.  INTRODUCTION\par}

A satisfactory theory that unifies Quantum Mechanics with General Relativity 
must start with a description of the fundamental degrees of freedom that are 
relevant at the Planck scale. Standard quantum field theories, including 
perturbative Quantum Gravity as well as supergravity theories, use as a 
starting point that Hilbert space can be thought of as spanned by the set of 
states of $N$ approximately freely moving physical particles: 
$$\HH\qu\bigg\{|\vec p_{(1)},\s_{(1)},\ \vec p_{(2)},\s_{(2)},\ \dots,\ \vec 
p_{(N)},\s_{(N)} \ket\ ,\qquad N=0,\,1,\,2,\,\dots\bigg\}\ ,\eqno(1.1)$$ 
where $\vec p_{(i)}$ are the momenta, and $\s_{(i)}$ discrete degrees of 
freedom such as spin and other quantum numbers.  Even though the particles 
may interact, one still describes a state of Hilbert space at any given 
instant of time using this language\fn{In non-Abelian gauge theories, there 
are further refinements: ghost particles have to be introduced, which play a role
in the renormalization procedure, but must be removed from the spectrum of
physical states.}. After constructing wave packets using 
these states as a basis, one can define the scattering matrix as the 
transformation matrix between ingoing and outgoing wave packets. 
 
For non-perturbative Quantum Gravity, this picture cannot be maintained.  Not 
only do trans-Planckian particles (elementary particles boosted towards 
energies beyond the Planck energy) cause non-renormalizable divergences in 
perturbation theory, but they also form mini-black holes, as soon as one 
would allow for the presence of several trans-Planckian particles moving in 
different directions. Therefore, the true Quantum Gravity Hilbert space must 
be very different from (1.1). 
 
Given a box with linear dimensions of the order of some length $R$. The 
highest energy state that can be described as being confined to this box must 
be a black hole with mass $M\approx R/Gc^2$. It is also the state with the 
largest entropy allowed inside the box. This observation dictates that the 
dimensionality of Hilbert space grows exponentially with the surface area of 
the box, as if we had a conventional field theory entirely restricted to the 
surface, with only fermionic basic fields and a cut-off at the Planck length. 
We refer to this doctrine as the `Holographic Principle'\ref1. 
 
On the one hand, the dimensionality of this Hilbert space is hardly big 
enough to describe all cis-Planckian states (states containing only particles 
with energies less than the Planck energy). On the other hand, we still would 
like to maintain Lorentz-invariance, which allows a single particle to be 
boosted to arbitrary energies.  The way out of this dilemma must be that, 
although single particle states allow to be Lorentz boosted to 
trans-Planckian energies, it will not be allowed to boost one particle in one 
direction and an other particle in an other direction to beyond the Planckian 
regime, without creating states that can also be described in a different 
way. If two trans-Planckian particles approach one another too closely, they 
should actually be described as being a black hole, and their combined state 
cannot be distinguished from black holes formed in some different way. 
 
Thus we see that any attempt at a description of Hilbert space that is as
sharply defined as possible, forces us to consider the presence of black
holes.  According to the equivalence principle in General Relativity, the
behavior of a single, large black hole can be derived from field theories in
flat space-time by means of a Rindler space transformation.  If this would
not be true at all for states in the immediate vicinity of a horizon, then it
would be difficult to imagine why this equivalence principle appears to work
flawlessly in ordinary General Relativity.  We consider it to be much more
likely that this principle can be replaced by some more complicated symmetry
transformation even in the full theory of Quantum Gravity.  The author has
advertised before, that one should be able to use this symmetry
transformation in the reverse direction, {\it i.e.\/}, it should be possible
to derive features of the Hilbert space of flat space-time by studying the
horizon of one black hole, preferrably of infinite size.  Surely, the Hilbert
space~(1.1) should be replaced by something else, but whatever it is, one
should still be able to define asymptotic wave packets of in- and out-states
-- it is alright if these contain trans-Planckian particles, as long as each
of these are very far separated from all the others.  Therefore, we do assume
that the notion of a scattering matrix still makes sense in Quantum
Gravity\ref2.

Although this idea was originally taken with much skepticism, it is now
generally being confirmed that also in terms of $D$-brane theories black
holes can be represented as pure quantum states\ref{3}.  However, insights
in the dynamics of a black hole horizon can be obtained without assuming the
validity of string theories or their relatives.  What one has to assume is
that not only quantum fields in the vicinity of a horizon, but also the black
hole as an entire unit should obey the laws of quantum mechanics, or, more
precisely, the processes of particle emission and absorption should be
described by a scattering matrix.  This is referred to as the $S$-matrix
Ansatz\ref2.

We insist that the $S$-matrix Ansatz should be appliccable first and foremost
for the prototype of all black holes, the Schwarzschild solution in 3+1
dimensions.  It is not difficult to argue that the $S$-matrix Ansatz may well
be a local property of any horizon, since the entropy of a black hole is
proportional to the horizon area.  Thus, the properties of Reissner-Nordstr\o
m and Kerr-Newmann black holes should follow directly from the pure
Schwarzschild case.  In contrast, extreme black holes will be much more
dubious objects.  Only string theorists appear to be able to imagine extreme
black holes as regular objects, but the physical black holes, among which the
astronomical ones, will never come close to the extreme case through either
accretion or decay.  No further attention will be given to extreme black
holes in this paper.

The mutual interactions between in- and outgoing particles are known, as soon 
as\fn{The reason for the factor $\scriptstyle 8\pi$ will 
become evident soon. As has been proposed earlier, we could use the symbol
$\scriptstyle\Gbar$ for $\scriptstyle 8\pi G$, and the associated units the
$\overline{\hbox{\medrm Planck}}$-units.} 
$$|g_{\m\n}(x)\, p^\m_\in p^\n_\out |\lap M_\Pl^2\ ,\qquad M_\Pl= 
\sqrt{\hbar c/8\pi G}\ ,\eqno(1.2)$$ 
where $p^\m_\in$ and $p^\n_\out$ are the momenta of an ingoing and an 
outgoing particle in a locally regular coordinate frame with metric\fn{Our 
signature convention for the metric $\scriptstyle g_{\m\n}$ is 
$\scriptstyle (-,+,+,+)$.} 
$g_{\m\n}(x)$, and we assume ordinary quantum field theory and perturbative 
quantum gravity to apply below Planckian energies.  However, since one is 
forced to limit oneself to the case that the in- and the outgoing particles 
stay separated by distances in the transverse direction that are large 
compared to the Planck length, we have not yet been able to cover all 
possible elements of Hilbert space this way, which may explain why as yet 
this approach did not naturally yield the Hawking entropy. 
 
The next step in the right direction could be made by replacing the algebra 
of known operators in Hilbert space by an obviously Lorentz covariant 
algebra\ref{4, 5, 6}, but then it was found that one is forced to choose either to 
have non-local commutators or incomplete commutators.  Choosing the latter 
option, we found that some of the horizon surface elements can be arranged 
according to a representation of the self-dual part of the Lorentz group 
$SO(3,1)$, but this representation is not unitary, and the commutators with 
the anti-self-dual parts are non-local.  This made the physical 
interpretation difficult, although one clearly finds hints of a space-time 
discreteness\ref7. 
 
In our approach, it is essential that the dimensionality of space-time is 
limited to the physical value 3+1.  Excursions to other dimensionalities may 
be useful exercises, but eventually we will have to deal with the physical 
case anyhow.  It was observed, however, that adding the electromagnetic force 
to the $S$-matrix Ansatz, essentially corresponds to the addition of a fifth, 
compactified Kaluza-Klein dimension, so it will certainly be of interest to 
study the $S$-matrix Ansatz also in 4+1 dimensions. 
 
Now this suggests that we can also venture towards the opposite direction. 
The proposal followed in this paper is, temporarily, to compactify also the 
third, physical, dimension.  This may amount simply to restricting oneself to 
`solutions' which are strictly periodic with respect to the $z$ coordinate. 
At a later stage, one could decide to lift this restriction, for instance by 
taking the limit where the period $R$ of the compactified coordinate tends to 
infinity.  Concretely, what we propose to do is to study the $S$-matrix 
Ansatz in 2+1 dimensions, adding as a refinement two conserved $U(1)$ 
charges, one for momentum conservation in the compactified dimension, and one 
for ordinary electric charge, and of course at a later stage we can 
accommodate for the other known interactions as well. 
 
The paper is organised as follows.  In Sect.  2, we summarise the derivation,  
by now standard, of the horizon algebra, first in 3+1 dimensions. Then we 
show how to derive a similar algebra for the case that one dimension is 
compactified. If, for the time being, we ignore the $U(1)$ charges altogether, the  
result is the simple commutator algebra of Eq.~(2.13), 
which is used as a starting point. Its representations are briefly described in 
Sect.~3. 
 Then, in Sect.~4, we describe the notion of a field operator in a
space-time point $(\tl x,\,t)$, in terms of the states in the horizon representation,
which illustrates the holographic principle.\ref{1}

In Sect.~5, we return to the question of decompactifying the compactified
dimension.  This, unfortunately, is not quite so straightforward, as the
results appears not to be Lorentz invariant, but the problem is seen to be
related to the fact that in more than 3 dimensions, the horizon has more than
one dimension, and this complicates the algebra.

\secbreak{\ni\bf 2.  THE HORIZON ALGEBRA}\par

An extensive discussion of the horizon algebra can be found in Ref\ref5.  We 
start by characterising ingoing states by their distribution of the ingoing 
momentum   
\hbox{$p_\in^-={1\over\sqrt2}(p^1-p^0)$} as a function of the transverse  
coordinates $\tl x=(y,z)$ of the 
impact point (as yet two-dimensional): 
$$|\in\ket\df|\{p_\in^-(\tl x)\}\ket\,.\eqno(2.1)$$  
Any change in the in-state of the form $\d p_\in^-(\tl x)$, produces a shift 
$\d x_\out^-$ among the outgoing particles, of the form $$\d x^-_\out(\tl 
x)=\int\dd\tl x'\,f(\tl x-\tl x')\d p_\in^-(\tl x')\,,\eqno(2.2)$$ 
where $f(\tl x)$ obeys the Poisson equation (if the background is flat)
$$\tl\pa^2f(\tl x)=-\d^2(\tl x)\,,\eqno(2.3)$$ 
in units where $8\pi G =1$. 
Representing the outgoing state by the distribution $p^+_\out(\tl x)$, one 
obtains the amplitude 
$$\bra \out | \in  \ket\,=\,\NN\exp\bigl[-i\int\dd^2\tl x\, 
\dd^2\tl x'p_\in^-(\tl x')f(\tl x-\tl x')p_\out^+ (\tl x)\bigr]\,.\eqno(2.4)$$ 
This allows us to identify the Hilbert space of outgoing states with that of the 
ingoing particles. Introducing operators 
$$\eqalign{p^-(\tl x)=p^-_\in(\tl x)\ ,&\qquad x^+(\tl x)\ ,\qquad 
[p^-(\tl x),\,x^+(\tl x')]=-i\d^2(\tl x-\tl x')\ ,\cr 
p^+(\tl x)=p^+_\out(\tl x)\ ,&\qquad x^-(\tl x)\ ,\qquad 
[p^+(\tl x),\,x^-(\tl x')]=-i\d^2(\tl x-\tl x')\ ,\cr}\eqno(2.5)$$ 
allows us to write 
$$p^-(\tl x)=-\tl\pa^2x^-(\tl x)\ ,\qquad p^+(\tl x)=\tl\pa^2x^+(\tl x)\ . 
\eqno(2.6)$$ 
 
In Ref\ref5, from Eq.~(15.11) onwards, it is attempted to rewrite these 
equations in a completely Lorentz-invariant way, leading to partly local 
commutation rules for the surface elements 
$$W^{\m\n}(\tl\s)=\e^{ab}\part{x^\m}{\s^a}\part{x^\n}{\s^b}\,.\eqno(2.7)$$ 
The route we will now take instead, is to choose one of the transverse  
coordinates,  $z=x_3$, to be compactified, 
having a period $R$. Here, the horizon is  taken to be running in the $(y,\,z)=(x_2,\,x_3)$
direction, the longitudinal coordinates being $x^1,\,x^0)$. 
We only insist that the equations~(2.5) and~(2.6) hold 
at transverse distances $|\d y|\gg R$, in which case the explicit  
$z$-dependence can be ignored. Taking in mind  that $p^\pm(\tl x)$ should be  
treated as distributions, and $x^\pm(\tl x)$ should be averaged over in the  
$z$ direction,  we define 
$$p^\pm(y)=\int\dd z\,p^\pm(y,z)\ ,\qquad x^\pm(y)={1\over R}\int\dd z\,x^\pm 
(y,z)\ .\eqno(2.7)$$ 
This turns Eqs~(2.5) and~(2.6) into 
$$[p^\pm(y),\,x^\mp(y')]=-i\d(y-y')\ ,\qquad p^\pm(y)=\pm R{\pa^2 
\over\pa y^2}x^\pm(y)\ .\eqno(2.8)$$ 
For the time being, we remove the factor $R$ here by replacing our units  
such a way that $$8\pi G/R=1\,.\eqno(2.9)$$ 
These are now the equations that hold for $|\d y|\gg 1$. From now on,  
in Eq.~(2.8), $R=1$.  
 
Invariance under the $SO(1,1)$ subgroup of the Lorentz group that corresponds to 
$(x,\,t)$ transformations, is obvious: 
$$\eqalign{p^+\ra A\,p^+\ ;&\qquad x^+\ra A\,x^+\ ;\cr 
p^-\ra A^{-1}p^-\ ;&\qquad x^-\ra A^{-1}x^-\ .\cr}\eqno(2.10)$$ 
Before restoring $SO(2,1)$ invariance, we rewrite our algebraic equations 
as follows: 
$$[\pa_y^2\,x^+(y),\,x^-(y')]=-i\d(y-y')\quad\ra\ \Big\{\matrix{[\pa_y x^+(y),\, 
\pa_y x^-(y')]&=&i\d(y-y')\,,\cr 
[\pa_y x^0(y),\,\pa_y x^1(y')]&=&i\d(y-y')\,.\cr}\eqno(2.11)$$ 
 
Replacing the $y$ coordinate by an arbitrary coordinate $\s(y)$, turns
this equation into
$$[\pa_\s x^0(\s),\,\pa_\s x^1(\s')]=i\pa_\s y(\s)\,\d(\s-\s')\,,\eqno(2.12)$$ 
where $\pa_\s y(\s)$ is required to reproduce the necessary Jacobian factor. 
Now we see that, writing $y=x_2(\s)$, allows us to rewrite Eq.~(2.12) as 
$$[\pa_\s x^\m(\s),\,\pa_\s x^\n(\s')]=i\e^{\m\n\l}\,g_{\l\k}\d(\s-\s')\pa_\s x^\k(\s)  
\ ,\eqno(2.13)$$ 
if $(\m,\,\n)=(0,\,1)$ or $(+,\,-)$.
The sign of the $2+1$ dimensional $\e$ symbol is set by 
$$\e^{012}=\e^{+-2}=+1\,.\eqno(2.14)$$
For $g_{\l\k}$ we take the flat metric  diag(-1,1,1). It is important to note that 
in the transformation from $y$ to $\s$, the Jacobian 
$\pa_\s y$ was assumed to be positive. This implies that the {\it orientation\/}
of the horizon, with respect to the observer, is of relevance to our considerations.
Note furthermore that the contribution of $\pa y/\pa\s$ helps to make the expression 
look Lorentz-covariant. 
Indeed, requiring invariance under the Lorentz group $SO(2,1)$, leads us to
expect Eq.~(2.13) to remain valid also when $\m$ or $\n$ reprsents the $y$ coordinate. 
This extremely non-trivial step hinges upon the assumption that the operators
$x^\pm(\tl x)$ in Eqs.~(2.5) indeed represent the position of the horizon
in the longitudinal coordinates, thus allowing us to view $(x^\pm,\,\tl x)$
as a Lorentz vector. The result of this assumption is that we obtained what we believe
to be the correct commutation rules for all coordinates, longitudinal and transverse
alike (though as yet reduced to 2+1 dimensions, see Sect.~5).

It is the new commutator~(2.13) that we will use as our starting point. 
In contrast to the general case in 3+1 dimensions, we are still dealing with 
entirely local commutators (that is, containing only Dirac delta functions 
in $\s$ and $\s'$), and in the next section, we exploit this observation. 
\secbreak 
\ni{\bf 3. REPRESENTATIONS} 
 
The commutator algebra (2.11) was derived directly from the $S$-matrix
Ansatz, and is valid when the transverse distance scales $\D y$ are large,
so that the transverse momenta $p_y$ are negligible.  This is important to
remember; in (2.11) the sideways shifts had been ignored.  The
representations of (2.11) are simply described by single functions,
$$|\{p^-(y)\}\ket\Or|\{p^-(y)\}\ket\Or|\{x^+(y)\}\ket\Or|\{x^-(y)\}\ket\ ,\eqno(3.1)$$
which can be transformed one into the other using (2.8).
 
The algebra (2.13) is assumed to have a wider validity; being Lorentz
invariant it handles the sideways shifts equally well as the longitudinal
shifts.  It replaces (2.11) when the transverse momenta also become large.

As Eq.~(2.13) contains derivatives with respect to the (arbitrary) coordinate 
$\s$, we are invited to integrate over $\s$. Let us divide the horizon into 
segments $A=[\s_1,\,\s_2]$, $B=[\s_2,\,\s_3]$, etc.. Define 
$$a_{(1)}^\m=x^\m(\s_2)-x^\m(\s_1)\ ;\qquad a_{(2)}^\m=x^\m(\s_3)-a^\m(\s_2)\ ; 
\qquad\hbox{etc.}\eqno(3.2)$$ 
Integrating Eq.~(2.13) over $\s$ yields the new commutation rules for the 
line segments $a_{(i)}^\m$ (see Fig.~1): 
$$[a_{(i)}^\m,\,a_{(j)}^\n]=i\d_{ij}\,\e^{\m\n\l}\,g_{\l\k}a_{(i)}^\k\,.\eqno(3.3)$$ 
Since the different segments commute, we can now concentrate on one segment 
$A$ only. As Eq.~(3.3) corresponds to the algebra of the Lorentz group 
$SO(2,1)$ (for each segment), its non-trivial representations are infinite dimensional. 
The Casimir operator is 
$$\left(\int_A\dd\s\,\pa_\s x^\m\right)^2={\bf a}^2=g_{\m\n}a^\m a^\n= 
{a^1}^2+{a^2}^2-{a^0}^2\,.\eqno(3.4)$$ 
This operator gives us some information about the separation between the end 
points of the segment $A$. If we fix it to some definite value, this corresponds 
to a gauge fixing condition for the coordinate $\s$, which, after all, had been 
chosen arbitrarily. Therefore, it is legitimate to give a constraint on the allowed 
value(s) for ${\bf a}^2$. We write 
$${\bf a}^2= -\ell(\ell+1)\,,\eqno(3.5)$$ 
although $\ell$ does not have to be an integer or half-integer. Special cases of 
importance will be the choices $\ell=\half$, $\ell=0$, and $\ell=-\half$.

\midinsert\cl{\epsffile{gthfig1.ps}}
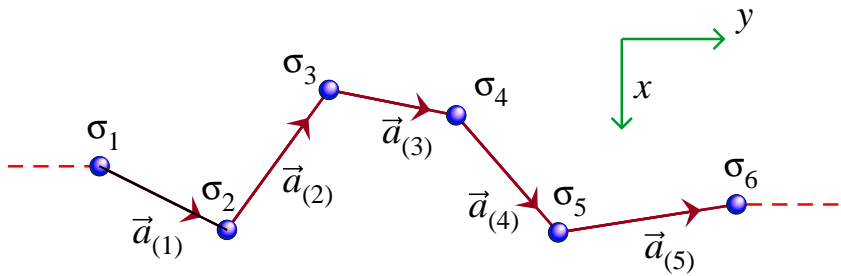
\cl{Fig. 1. Segmentation of the horizon.}\endinsert 
 
The representations are constructed by diagonalising $a^0=m$, and using the 
raising and lowering operators $a^\pm$: 
$$\eqalignno{a^\pm=a^1\pm ia^2\ ,  \qquad &[a^0,\,a^\pm]=\pm a^\pm\,,&(3.6)\cr 
 a^\mp a^\pm={\bf a}&^2+a^0(a^0\pm1)\,. &(3.7)}$$ 
{}From this: 
$$a^\pm|m,\,\ell\ket=\sqrt{m(m\pm1)-\ell(\ell+1)}\,|m\pm1,\,\ell 
\ket\,.\eqno(3.8)$$ 
If $\ell$ is real and positive, the values of $m$ may be taken to be 
$$m=\ell+1,\,\ell+2,\dots\,,\,\infty\,,\qquad\hbox{or}\qquad 
m=-\ell-1,\,-\ell-2,\dots\,,\,-\infty\,.\eqno(3.9)$$ 
In addition, for all real values of ${\bf a}^2$, with $\ell$ either real or complex, 
one may have series of $m$ values that are unbounded above and below: 
$$m=m_0,\ m_0\pm1,\ m_0\pm2,\,\dots\ \pm\infty\,.\eqno(3.10)$$ 
 
The representations (3.9) will be called timelike. Even though  ${\bf a}^2$  
may be positive \hbox{(${\bf a}^2<\fract14$)}, the vector $\bf a$ may be seen to 
lie either on the positive or on the negative wing of a ``timelike'' 
hyperbola. The series~(3.10) will be called spacelike, although one could 
also admit negative values for ${\bf a}^2$ here. 
 
A special case is $\ell=-\half$, which we will call lightlike. In this case, 
the square root in Eq.~(3.8) can be drawn: 
$$a^\pm|m,-\half\ket=(m\pm\half)\,|m\pm1,-\half\ket\,.\eqno(3.11)$$ For 
negative $m$ values, this implies a minor revision of the sign conventions 
for the states $|m,-\half\ket$. 
 
We conclude that the representations of our horizon algebra are characterised 
by a number $\ell$ that is either real or of the form 
$\ell=-\half+i\,\hbox{Im}(\ell)$, and that there is a special case 
$\ell=-\half$, where the explicit expressions for the operators simplify. 
The full Hilbert space $\HH$ for the representations of the algebra~(2.13) is 
the product of the hilbert spaces $\HH_A$, $\HH_B$, \dots of the (connected) 
line segments $A$, $B$, etc. The vectors ${\bf a}_{(1)}$, ${\bf a}_{(2)}$, 
\dots, can be added to describe larger horizon segments, in a way similar to 
the procedure of adding angular momenta, although, of course, there are fewer 
selection rules, due to the fact that the vectors ${\bf a}_{(i)}$ are Lorentz 
vectors. 
 
The number of states is always infinite, but if 
$\big({a^1}^2+{a^2}^2\big)^{1/2}$ can be taken to be a measure for the ``length'' of a 
section of the horizon, and if we look at all states where the total length 
of the horizon is limited to some value, we find that the number of allowed 
values for $\{a^0_{(i)}\}$ is bounded after all, and it will naturally 
diverge as the exponential of the horizon area (= length), as one should 
expect from the entropy formula. 
 
\secbreak 
\ni{\bf 4. HOLOGRAPHIC FIELDS} 
 
As it was emphasized in the Introduction, and in earlier papers, this Hilbert 
space not only describes the horizon of a black hole, but it should also 
describe the Hilbert space of the entire universe, since the Rindler space 
transformation allows us to relate horizon dynamics with the dynamics of flat 
space-time. Since we have chosen the metric $g_{\m\n}$ to be that of flat 
space-time, the black hole under considerartion is of infinite size, and 
therefore all points of space-time that are at a finite metric distance from 
our horizon, belong to the black hole's environment. All field operators 
$\f(\tl x,t)$, for finite values of $\tl x$ and $t$, should operate within 
the same Hilbert space.  

How to define a field operator in our Hilbert space, be it (3.1) or the
representation of (2.13) as derived in Sect.~3, is not at all evident.
A scalar field (in 2+1 dimensions) would normally be defined as\fn{Sign conventions
in the exponent are here chosen such that $\pa\f/\pa t=-i[H,\f]$.}
$$\f(\tl x,t)=\int{\dd^2\tl k\over\sqrt{2k^0(\tl k)(2\pi)^2}}\Big(
a(\tl k)e^{-i\tl k\cdot\tl x+ik^0t}+a^\dagger(\tl k)e^{i\tl k\cdot\tl x-
ik^0 t}\Big)\,,\eqno(4.1)$$
where $a$ and $a^\dagger$ annihilate and create exactly one particle.
Concentrating first on the algebra (2.11) with representation (3.1), we
write $$\f(\tl x,t)=\sqrt{(2\pi)^{-2}}\int\dd^2\tl k \hat\f(\tl k,t)
\,e^{i\tl k\cdot\tl x}\,,\eqno(4.2)$$
and we may impose that the sideways momenta $p_y$ and the mass $\m$ of
the particles involved are negligible\fn{In this section, we use the notation
$\tl x=(x,\,y)$ and $\tl k=(k_x,\,k_y)$.}.
In that case, the equation of motion for the field reads
$$(k^+k^-+k_y^2+\m^2)\hat\f(\tl k,t)\ \approx\  k^+k^-\hat\f(\tl k,t)\ 
\approx\ 0\,,\eqno(4.3)$$
which implies that either $k^+$ or $k^-$ vanish; our field then consists
of an ingoing component $\f_\in$ and an outgoing component $\f_\out$.
Since, in this case, $k^0$ only depends on $k_x$ and not on $k_y$, we may
introduce the mixed Fourier transform, 
$$\f_\in(\tl x,t)=\sqrt{(2\pi)^{-1}}\int\dd k_x\,e^{ik_x(x+t)}\check\f_\in
(k_x,y)\,,\eqno(4.4)$$ to obtain
$$ \check\f_\in(k_x,y')\Big|\,p^-(y)\Big\ket= \sqrt{N\over 2|k_x|}\,\Big|\,
p^-(y)+\d(y-y')k_x\sqrt{2}\,\Big\ket\,,\eqno(4.5)$$
where $N$ refers to the number of identical particles already present in the ket-state
in case of annihilation, or in the bra state in case of creation. We used
$k^0=|k_x|$. The factor $\sqrt2$ is a normalization factor.

Temporarily omitting the square root in Eq.~(4.5), we find that the field
operator can be written as
$$\check\f_\in(k_x,y')=C\exp\Big(-ix^+(y')\,k_x\sqrt2\Big)\,,\eqno(4.6)$$
since $-x^+(y')$ is precisely the operator that produces the required shift
in the function $p^-(y)$, according to (2.5). Using (4.4),
$$\eqalign{\f_\in(x^+,x^-,y')&\,=\,{C\over\sqrt{2\pi}}\int\dd k_x e^{i\sqrt2
\big(k_xx^+-k_xx^+(y')\big)} \cr&\,=\,C'\d\left(x^+-x^+(y)\right)\,,}\eqno(4.7)$$
which is independent of $x^-$ just because it is an ingoing field. $C'$ is
just a new constant.
 
In order to express the action of this operator in terms of operators of
the algebra (2.13) or (3.2), we convert to the $\s$ coordinates:
$$\f_\in(x^+,x^-,y')\approx C'\int\dd\s\left|\part y\s\right|\, \d\big(y(\s)-y'\big)\,\d\big(
x^+(\s)-x^+\big)\,.\eqno(4.8)$$
Similarly,
$$\f_\out(x^+,x^-,y')\approx C'\int\dd\s\left|\part y\s\right|\, \d\big((y(\s)-y'\big)\,\d\big(
x^-(\s)-x^-\big)\,,\eqno(4.9)$$
and the original field $\f(\tl x,t)$ could be defined as $$\f(\tl x,t)=
\f_\in(x^+,y)+\f_\out(x^-,y)\,.\eqno(4.10)$$ 

In these expressions, we used the `approximate' symbol ($\approx$) because the
expressions were derived for the case that they act on states with smooth
$y$ dependences (small values for $p_y$). The exact expressions are not very
useful for fields since the operators $x^+(\s)$ and $y(\s)$ do not commute,
so that the order of the Dirac deltas may not be altered, and hence the
transformation properties of these fields under rotations an Lorentz transformations
would be problematic. 

Yet, it is of importance to try to define as precisely as possible field operators
with local commutation rules. Eq.~(4.10) may be required to hold only in the
limit of vanishing $p_y(\s)$. It is not clear to the author what the best
possible procedure for this is. Our approximate expressions (4.8)--(4.10)
suggest for instance to define
$$\f(x,y,t)=C''\sum_{\hbox{\smallrm orderings}}\int\dd\s\,\d\big(x^1(\s)-x\big)
\d\big(x^2(\s)-y\big)\d\big(x^0(\s)-t\big)\,,\eqno(4.11)$$
where the delta functions must be in all possible orders, and $C''$ is again
a new numerical constant. The equations of motion for such fields, and
{}from there the particle content of the theory, should
follow from these expressions. One thing appears to be obvious\ref5: the time coordinate
$t$ tends to be quantized in units $ 8\pi G\hbar/c^5 R$. 
The fact that the time quantum
is inversely proportional to $R$ should not come as a surprise. The maximal energy
admitted in a compactified world is linearly proportional to $R$.

	\secbreak\ni{\bf 5. A NOTE ON DECOMPACITFICATION}\medskip
In our compactified world, the momenta in the compactified dimension
$z$ in all respects act as new electric charges. Electric charge
has been considered in Refs.\ref{2, 5}, and by using these results, incorporation of these
compactified dimensions should be straightforward. It was found that
electric charge dnsity operators $\r_\in(y)$ and $\r_\out(y)$ must be 
defined separately for the in-Hilbet space and for the out-Hilbert space.
Similarly, we have the phase operators $\f_\in(y)$ and $\f_\out(y)$,
living in the in- and the out-Hilbert space, respectively (strictly
speaking, these objects are only defined {\it modulo\/} their
period $2\pi$, so that they are not good as operators, but we can give
unambiguous meaning to their derivatives, $\pa_y\r_{\in,\,\out}(y)$).

Correspondingly, this leads to introducing (de-)compactified coordinate 
operators $a^i_\in(y)$ and $a^i_\out(y)$ ($i=3$ or more), obeying the commutation rule
$$[\pa_ya^i_\in(y),\, \pa_ya^j_\out(y')]= ig^{ij}\d(y-y')\,,\eqno(5.1)$$
which, in $\s$ coordinates would read
 $$[\pa_\s a^i_\in(\s),\, \pa_\s a^j_\out(\s')]= ig^{ij}(\s)\pa_\s y\,\d(y-y')\,,\eqno(5.2)$$
and $a^i,\ i=3,\dots$ is expected to commute with $a^\m,\ \m=0,1,2.$.

It is difficult to see how to turn these equations into Lorentz covariant ones.
The reason for this is, presumably, that the horizon should not be taken to
depend on a single coordinate $\s$ but on two (or more, if space-time
has more than 4 dimensions) coordinates $\tl\s=(\s^1,\,\s^2)$.
For the correct, Lorentz covariant algebra, we have to return to the
original treatment of Ref\ref{4, 5}. 

Alternatively, there is the following possibility to be considered. We could
rewrite the algebra (3.3) in terms of the operators (temporarily suppressing the indices
$i$),
$$T_{\m\n}=\e_{\m\n\l}a^\l\,,\eqno(5.3)$$
which then obey$$[T_{\m\n},\,T_{\k\l}]=g_{\m\k}T_{\n\l}-g_{\m\l}T_{\n\k}-g_{\n\k}T_{\m\l}
+g_{\n\l}T_{\m\k}\,.\eqno(5.4)$$
It is tempting to generalize this commutator to larger numbers of dimensions.
This would slightly deviate from what one can derive directly in 4-dimensional
spacetime\ref5, but it is conceivable that a consistent theory can be obtained
along these lines.

Our theory did result in a scheme where Hilbert space is represented by
integers, and if the length of the horizon, to be defined as $\sum_i\big({a^1_i}^2
+{a^2_i}^2\big)^{1/2}$, is kept finite, there are only a finite number of states. Yet there
is Lorentz-invariance. This must imply that the transPlanckian particle contents
os a state are not freely adjustable, but follow from the configurations of the 
cis-Planckian particles present.

\secbreak
{\bf\ni References}\medskip
 
\item{1.} G. 't Hooft, {\it Dimensional Reduction in Quantum Gravity},
in {\it Salamfestschrift: a
collection of talks}, World Scientific Series in 20th Century Physics, vol.
4, ed. A. Ali, J. Ellis and S. Randjbar-Daemi (World Scientific, 1993),
THU-93/26, gr-qc/9310026;
G.~'t Hooft, {\it Black holes and the dimensionality of space-time},
Proceedings of the Symposium ``The Oskar Klein Centenary'', 19-21 
Sept. 1994, Stockholm, Sweden. Ed. U.~Lindstr\"om, World Scientific 1995, p. 122;
L. Susskind, J. Math. Phys. {\bf 36} (1995) 6377.
 
\item{2.} G. 't Hooft, Physica Scripta, Vol. {\bf T15} (1987) 143; {\it id.},
 Nucl. Phys. {\bf B335} (1990) 138.
 
\item{3.} See e.g. J.M. Maldacena, {\it Black holes in string theory}. Princeton University
 Dissertation, June 1996, hep-th/9607235.

\item{4.} G. 't Hooft, {\it in} CCAST/WL Meeting on Fields, Strings and 
Quantum Gravity, Beijing, 1989 (Gordon and Breach, London).

\item{5.} G. 't Hooft, J. Mod. Phys. {\bf A11} (1996) 4623, gr-qc/9607022. 

\item{6.} 
S. de Haro, {\it Class. Quant. Grav. \bf 15} (1998) 519.
 
\item{7.}  G. 't Hooft, in
Proc. of NATO Advanced Study Institute on 
Quantum Fields and Quantum Space Time, Carg\`ese, July 22-August 3, 1996.
Eds.: G. 't Hooft et al., Plenum, New York. NATO ASI Series B364 
(1997) 151 (gr-qc/9608037); 
 G. 't Hooft, Class. Quantum Grav. {\bf 13}  1023.

\end